\documentclass[5p,times,twocolumn]{elsarticle}
\usepackage{lineno,hyperref}
\usepackage{subcaption}
\usepackage{tikz}
\usepackage{pgfplotstable}
\usepackage{pgfplots}
\usepackage{multirow}
\usepackage{float}
\usepackage{textcomp}
\usepackage{amsmath,amsfonts,amssymb,bbm}	
\usepackage{multicol}

\usepackage[section]{placeins}

\modulolinenumbers[5]

\biboptions{authoryear}

\usepackage{booktabs} 
\usepackage{graphicx}
\usepackage{lscape} 
\usepackage{placeins} 
\usepackage{hyperref}
\usepackage{rotating} 
\usepackage{siunitx}

\usepackage{blindtext}

\begin{document}

\begin{frontmatter}

\title{DDoS-UNet: Incorporating temporal information using Dynamic Dual-channel UNet for enhancing super-resolution of dynamic MRI} 

\author[1,2,3,4]{Soumick Chatterjee\corref{equalcontribution}}
\author[3,4,5]{Chompunuch Sarasaen\corref{equalcontribution}}
\author[4,5,7]{Georg Rose}
\author[1,2,7]{Andreas N{\"u}rnberger}
\author[3,4,6,7,8]{Oliver~Speck}

\cortext[equalcontribution]{S. Chatterjee and C. Sarasaen have Equal Contribution}

\address[1]{Faculty of Computer Science, Otto von Guericke University Magdeburg, Germany}
\address[2]{Data and Knowledge Engineering Group, Otto von Guericke University Magdeburg, Germany}
\address[3]{Biomedical Magnetic Resonance, Otto von Guericke University Magdeburg, Germany}
\address[4]{Research Campus STIMULATE, Otto von Guericke University Magdeburg, Germany}
\address[5]{Institute for Medical Engineering, Otto von Guericke University Magdeburg, Germany}
\address[6]{German Center for Neurodegenerative Disease, Magdeburg, Germany}
\address[7]{Center for Behavioral Brain Sciences, Magdeburg, Germany}
\address[8]{Leibniz Institute for Neurobiology, Magdeburg, Germany}

\begin{abstract}
Magnetic resonance imaging (MRI) provides high spatial resolution and excellent soft-tissue contrast without using harmful ionising radiation. Dynamic MRI is an essential tool for interventions to visualise movements or changes of the target organ. However, such MRI acquisition with high temporal resolution suffers from limited spatial resolution - also known as the spatio-temporal trade-off of dynamic MRI. Several approaches, including deep learning based super-resolution approaches, have been proposed to mitigate this trade-off. Nevertheless, such an approach typically aims to super-resolve each time-point separately, treating them as individual volumes. This research addresses the problem by creating a deep learning model which attempts to learn both spatial and temporal relationships. A modified 3D UNet model, DDoS-UNet, is proposed - which takes the low-resolution volume of the current time-point along with a prior image volume. Initially, the network is supplied with a static high-resolution planning scan as the prior image along with the low-resolution input to super-resolve the first time-point. Then it continues step-wise by using the super-resolved time-points as the prior image while super-resolving the subsequent time-points. The model performance was tested with 3D dynamic data that was undersampled to different in-plane levels. The proposed network achieved an average SSIM value of 0.951±0.017 while reconstructing the lowest resolution data (i.e. only 4\% of the k-space acquired) - which could result in a theoretical acceleration factor of 25. The proposed approach can be used to reduce the required scan-time while achieving high spatial resolution - consequently alleviating the spatio-temporal trade-off of dynamic MRI, by incorporating prior knowledge of spatio-temporal information from the available high resolution planning scan and the existing temporal redundancy of time-series images into the network model.
\end{abstract}


\end{frontmatter}



\section{Introduction} 
\label{sec:introduction}
Magnetic resonance imaging (MRI) does not rely on ionising radiation and can provide high spatial resolution with superior visualisation of soft-tissue contrast. MR images can also offer better differentiation between fat, water and muscle than other imaging modalities. Therefore, image guidance based on MRI is a favourable tool for identifying and characterising tumours in interventions~\citep{barkhausen2017white,mahnken2009ct}. Interventional applications in real-time or near real-time, such as MR-guided liver biopsy, show excellent contrast between the target organ or structure and adjacent soft tissue while visualising the changes of internal organs during an examination. In such applications, dynamic MRI is used, which is obtained by acquiring the k-space data (in frequency domain) continuously and reconstructing a sequence of images over time~\citep{bernstein2004handbook}. However, while achieving high temporal resolution, these acquisitions suffer from the restricted spatial resolution because only a limited part of the data can be measured (undersampling). Consequently, the resultant image might have reconstruction artefacts due to the violation of the Nyquist criterion~\citep{shannon1949communication}, and also leads to image resolution loss. This is known as the spatio-temporal trade-off of dynamic MRI and has been demonstrated as one of the main research problems  ~\citep{lustig2006kt,lustig2007sparse,jung2009k,zhang2010magnetic}. Although common approaches such as compressed sensing ~ \citep{lustig2007sparse} can utilise the spatial and temporal correlation of the data to accelerate the data acquisition, the iterative processes could hinder real-time applications such as intervention MRI. 


Super-resolution (SR) is a process of estimating a high-resolution image from a low-resolution counterpart. Several deep learning based super-resolution algorithms have been proposed ~\citep{dong2015image,lim2017enhanced,ledig2017photo,zeng2018simultaneous,he2020super}. The existing SR techniques can be categorised into two major groups: single image super-resolution (SISR) and video super-resolution (VSR). In contrast to SISR, VSR exploits the temporal information in a sequence of images to enhance the spatial resolution and frame rate~\citep{yamaguchi2010video,caballero2017real,lucas2019generative}. Besides, some literature investigated the use of temporal information incorporation and reported its potential for improving the image quality of dynamic MRI reconstruction~\citep{rasch2018dynamic,kofler2019spatio,kustner2020cinenet}.

To further improve the super-resolved image quality, additional prior information had been integrated into the super-resolution process ~\citep{segall2004bayesian,belekos2010maximum}. The prior information can be incorporated in multi-channel training to enhance the results~\citep{chatterjee2020retrospective}. A multi-channel network allows better feature extractions when learning with multiple types of channels~\citep{araki2015exploring}. Multi-channel training has been used across numerous applications including image recognition~\citep{barros2014multichannel}, speech recognition~\citep{wang2018multi,nugraha2016multichannel}, audio classification~\citep{casebeer2019multi}, natural language processing ~\citep{xu2019modeling}, etc. This paper extends the previous work into the temporal domain~\citep{sarasaen2021fine} by exploiting dual-channel inputs (prior-image and low-resolution image) in the deep learning model - to learn the temporal relationship between time-points, while also learning the spatial relationship between low- and high-resolution images, to perform SISR, using the proposed DDoS (\textbf{D}ynamic \textbf{D}ual-channel \textbf{o}f \textbf{S}uper-resolution) approach. 



\subsection{Related Work}\label{sec:related}
The UNet architecture~\citep{Ronneberger2015}, including its 3D version~\citep{cciccek20163d}, is a versatile neural network consisting of two paths: contraction and expansion. Originally proposed for image segmentation, different flavours of UNet have been developed and deployed in plenty of applications such as image segmentation~\citep{milletari2016v,zhou2018unet++,oktay2018attention,chatterjee2020ds6}, audio source separation~\citep{jansson2017singing,stoller2018wave,choi2019phase} and image reconstruction~\citep{hyun2018deep,iqbal2019super}. 3D UNet and its variants have been used for MR super-resolution as well~\citep{pham2019multiscale,sarasaen2021fine,chatterjee2021shuffleunet}. Furthermore, UNet has been extended to multi-channel and dual-branch to incorporate prior-information~\citep{chatterjee2020retrospective}.

Since medical images are mainly used for diagnosis, evaluation using perception-based metrics are more suitable than pixel-wise metrics. Perceptual loss ~\citep{johnson2016perceptual} has demonstrated the ability to improve image quality perceptually, yielding superior results and reducing blurriness than classical pixel-based metrics such as L1 or L2~\citep{gatys2016imagePlossStyleTransf,ghodrati2019mrPLoss}. A recent study from ~\cite{zhang2018unreasonable} presented that deep feature extractions, which were obtained from the trained network, could be utilised to deal with excessive blurry images and showed that perceptual similarity is the rising property that has been shared among deep visual representations. Previous work~\citep{sarasaen2021fine} has also demonstrated the potential of applying a perceptual loss network to improve the results of image super-resolution. 




\subsection{Contributions}\label{sec:contrib}
This paper extends the research of Single-Image Super-Resolution (SISR) of dynamic MRIs treating each time-point as individual 3D volumes, by incorporating the temporal information into the network model using the proposed \textbf{DDoS-UNet} framework. The proposed method super-resolves the low-resolution dynamic MRI with the help of a static prior scan and by exploiting the temporal relationship between the different time-points. The method has been evaluated using Cartesian undersampling by taking different amounts of the centre k-space data, up to a theoretical acceleration factor of 25. 


\section{Methodology}\label{sec:Methodology}
In this work, the dynamic training data was initially generated from the benchmark dataset due to the lack of dynamic abdominal data. After that, it was undersampled, and a modified UNet model was trained on that. The dual-channel input consists of the low-resolution image of the current time-point ($LR_TPn$) and the super-resolved image of the previous time-point ($HR_TPn-1$). The network was trained and tested with different levels of undersampling. 


\subsection{Super-resolution Reconstruction}\label{sec:srrecon}
The reconstruction of the high-resolution image from the corresponding low-resolution image can be modelled as: 
\begin{equation}
    \hat{I}_{HR} =  \digamma(I_{LR}; \theta)
    \label{eqn:SR} 
\end{equation}

where $I_{LR}$ is the low-resolution image, $\hat{I}_{HR}$ is the super-resolved image, $\digamma$ is the mapping function which models the spatial super-resolution relationship between the corresponding low- and high-resolution images using a given set of parameters $\theta$~\citep{wang2020deep}. The SR image reconstruction is an ill-posed problem to approximate the super-resolved image from a given low-resolution counterpart, where the network model is trained to optimise the objective function: 


\begin{equation}
    \begin{split} \hat{\theta} = {arg\,min} 
    {\mathcal{L}(\hat{I}_{HR},I_{HR})+\lambda R(\theta)} \end{split}
    \label{eqn:SR_loss}
\end{equation}

The operator $\mathcal{L}(\hat{I}_{HR},I_{HR})$ describes the loss function between the predicted super-resolved image $\hat{I}_{HR}$ and a ground-truth image $I_{HR}$, $R(\theta)$ is a regularisation term and $\lambda$ denotes the regularisation parameter~\citep{sarasaen2021fine}.


\subsection{Network Architecture} \label{sec:proposedArchitecture}
\begin{figure*}[!htbp]
\centering
\includegraphics[width=\textwidth]{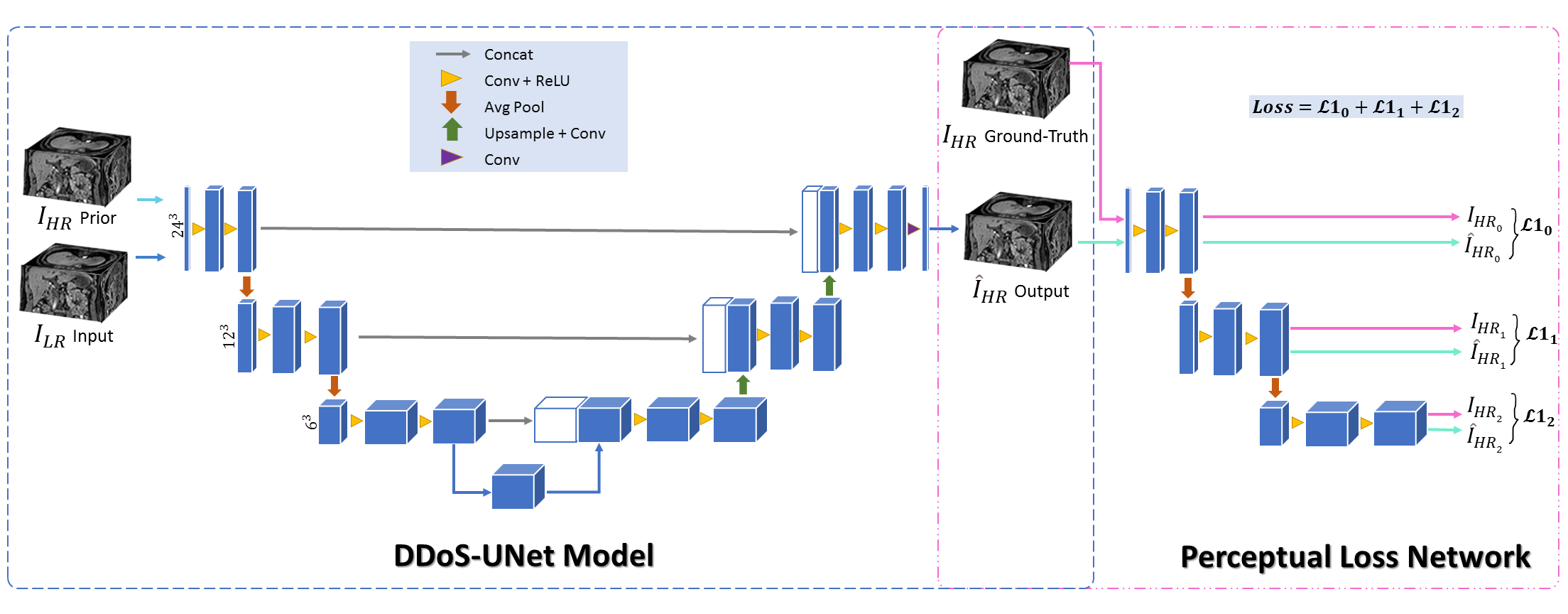}
\captionsetup{justification=centering}
\caption{Network Architecture.}
\label{fig:DDoS}
\end{figure*}

The 3D UNet architecture from the previous work~\citep{sarasaen2021fine} was extended using multi-channel for supplying prior-information~\citep{chatterjee2020retrospective} to create the proposed \textbf{D}ynamic \textbf{D}ual-channel \textbf{o}f \textbf{S}uper-resolution UNet architecture (DDoS-UNet, or simply DDoS), as shown in Fig.~\ref{fig:DDoS}. The basic architecture of the UNet is similar to the previous work~\citep{sarasaen2021fine} - except for two differences, having contracting (encoding) and expanding (decoding) paths. The contracting path is made of three blocks, each of the blocks comprises of two pairs of 3D convolutional layers (kernel size:3, stride:1, padding:1) and ReLU activation functions, followed by average pool layers (kernel size: 2) - making the output size of the block half the size of the input received by that block. The expanding path also consists of three blocks, each consisting of a pair of trilinear upsampling layer (scale factor:2) and 3D convolutional layer (kernel size:1, stride:1, padding:0), unlike the original work which used 3D convolutional transpose layers (first difference with the earlier model); followed by a convolutional block similar to the contracting path, except for the pooling layers. It is noteworthy that initial experiments were performed using 3D convolutional transpose layers similar to the earlier model, but for volumetric super-resolution this model resulted in checkerboard artefacts~\citep{odena2016deconvolution}. This can be attributed to the fact that overlapped portions of the patches are averaged in the patch-based super-resolution - mitigating the checkerboard problem, but in volumetric super-resolution, there is no averaging operation that could mitigate this effect. Each block of the expanding path increases the size of its input by a factor of two. Inside these expanding path blocks, after upsampling the input using trilinear-convolution pair, the output is concatenated with the input coming from a similar depth of the contraction path - known as skip-connections. The initial layer of the network provides an output of 64 feature maps. Then, each block of the contraction path increases the number of feature maps by two, whereas each of the expanding path blocks decreases it by two. Finally, a fully connected 3D convolutional layer (kernel size: 1, stride: 1, padding: 0) is applied to merge all the feature maps to generate the final output. The other difference between the earlier UNet~\citep{sarasaen2021fine} and this DDoS-UNet is the fact that the initial layer of the network receives two input channels rather than one. 

Since the UNet-like architectures warrant for the matrix size of the input to be the same as the output (ground-truth), the low-resolution input volumes were interpolated using trilinear interpolation with the interpolation factor equivalent to the acceleration factor before providing them as input to the DDoS-UNet model.  


\subsubsection{DDoS: Working Mechanism and Theory} \label{sec:ddos}
The DDoS-UNet works with dynamic MRIs while using the static planning scan as a prior image. Initially, the network is supplied with a patient-specific fully sampled high-resolution (HR) static prior scan on the first channel and the first time-point (TP0) of the undersampled low-resolution (LR) dynamic MRI on the second channel. It is to be noted that the static planning scan is acquired with the same protocol as the dynamic scan, but they are not co-registered. Given this pair of HR-LR images, DDoS-UNet super-resolves the LR to obtain the TP0 of the super-resolved (SR) HR dynamic MRI. This initial phase is termed here as the "Antipasto" phase as it precedes the main reconstruction phase. The reconstruction phase starts by supplying this SR-TP0 on the first channel, while the LR-TP1 is supplied on the second channel of the network to generate SR-TP1. This process is continued recursively for all the subsequent time-points. This can be formulated by modifying Eq.~\ref{eqn:SR} as : 
\begin{equation}
    \hat{HR}_{TP(n)} =  \digamma(LR_{TP(n)}; \hat{HR}_{TP(n-1)}; \theta)
    \label{eqn:DDoS} 
\end{equation}
where $\hat{HR}_{TP(n)}$ is the super-resolved time-point, $LR_{TP(n)}$ is the low-resolution time-point, $\hat{HR}_{TP(n-1)}$ is the super-resolved previous time-point, $\digamma$ is the super-resolution model that maps those three images, and $\theta$ is the set of parameters of $\digamma$. The authors hypothesise that the network learns two different representations: the temporal relationship between $\hat{HR}_{TP(n)}$ and $\hat{HR}_{TP(n-1)}$, the super-resolution relationship between $LR_{TP(n)}$ and $\hat{HR}_{TP(n)}$. If $\Psi$ is the DDoS operator and $\Phi$ is the set of parameters of the DDoS network, this hypothesis can be formulated as:
\begin{equation}
    \Psi(\Phi) =  \digamma_1(LR_{TP(n)}; \hat{HR}_{TP(n)}; \theta_1) + \digamma_2(\hat{HR}_{TP(n-1)}; \hat{HR}_{TP(n)}; \theta_2)
    \label{eqn:DDoS_param} 
\end{equation}
where $\digamma_1$ is the super-resolution operator learnt for the relationship between $LR_{TP(n)}$ and $\hat{HR}_{TP(n)}$, whereas $\theta_1$ is its parameters;  $\digamma_2$ is the temporal-relationship operator learnt for the relationship between $\hat{HR}_{TP(n-1)}$ and $\hat{HR}_{TP(n)}$, where $\theta_2$ is its parameters. 

It is worth mentioning that the patch-based super-resolution idea of the previous work~\citep{sarasaen2021fine} was dropped in this current research due to the working theory of DDoS-UNet. Due to physiological movements, the organs can move in and out of the $24^3$ patches (as used in the previous work). Consequently, the supplied $LR_{TP(n)}$ and $\hat{HR}_{TP(n-1)}$ patches might not contain similar organs - making the hypothesis of the temporal-relationship operator $\digamma_2$ of Eq.~\ref{eqn:DDoS} invalid. Hence, this work performs volumetric super-resolution (using complete 3D volumes) instead of 3D patch-based super-resolution. 

\subsection{Data} \label{sec:data}
The proposed method was trained using the publicly-available abdominal benchmark dataset: the CHAOS dataset (T1-dual images, in- and opposed phase)~\citep{kavur2021chaos}, comprising 80 volumes (40 subjects, in-phase and opposed-phase for each subject). Dynamic training data was generated artificially by applying random elastic deformation, explained in detail in Sec.~\ref{sec:chaosdyn}. The dataset was divided into training and validation sets with a ratio of 70:30. For testing the approach, high-resolution 3D static (breath-hold) and 3D "pseudo"-dynamic (free-breathing) scans for 25 time-points of five healthy subjects were acquired using a 3T MRI (Siemens Magnetom Skyra). Each subject's static and dynamic scans were acquired in different sessions using the same sequence, parameters, and volume coverage. All the datasets (except the high-resolution static scans) were artificially undersampled to simulate the low-resolution datasets. The acquisition parameters of the datasets are listed in Table~\ref{tab:MRIparams_CHAOS_3DDyn}. 

\begin{table*}
\centering
\caption{MRI acquisition parameters CHAOS dataset and subject-wise 3D dynamic scans. Static scans were performed using the same subject-wise sequence parameters as the dynamic scans for one time-point (TP), acquired at a different session.}
\label{tab:MRIparams_CHAOS_3DDyn} 
\resizebox{\textwidth}{!}{
\begin{tabular}{llllll}
\toprule
                            & \begin{tabular}[c]{@{}l@{}}CHAOS\\ (40 Subjects)\end{tabular} & \begin{tabular}[c]{@{}l@{}}Protocol 1\\ (2 Subjects)\end{tabular} & \begin{tabular}[c]{@{}l@{}}Protocol 2\\ (1 Subject)\end{tabular} & \begin{tabular}[c]{@{}l@{}}Protocol 3\\ (1 Subject)\end{tabular} & \begin{tabular}[c]{@{}l@{}}Protocol 4\\ (1 Subject)\end{tabular}\\ \toprule
Sequence &
  \begin{tabular}[c]{@{}l@{}}T1 Dual In-Phase \\ \& Opposed-Phase\end{tabular} &
  T1w Flash 3D &
  T1w Flash 3D &
  T1w Flash 3D &
  T1w Flash 3D \\ \midrule
Resolution &
  \begin{tabular}[c]{@{}l@{}}1.44 x 1.44 x 5 - \\ 2.03 x 2.03 x 8 $mm^3$\end{tabular} &
  0.90 x 0.90 x 4 $mm^3$  &
  0.90 x 0.90 x 4 $mm^3$ &
  0.90 x 0.90 x 4 $mm^3$ &
  1.00 x 1.00 x 4 $mm^3$ \\ \midrule
FOV x, y, z &
  \begin{tabular}[c]{@{}l@{}}315 x 315 x 240 - \\ 520 x 520 x 280 $mm^3$\end{tabular} &
  300 x 225 x 176 $mm^3$ &
  350 x 262 x 176 $mm^3$ &
  350 x 262 x 192 $mm^3$ &
  350 x 262 x 176 $mm^3$ \\ \midrule
Encoding matrix &
  \begin{tabular}[c]{@{}l@{}}256 x 256 x 26 - \\ 400 x 400 x 50\end{tabular} &
  320 x 240 x 44 &
  384 x 288 x 44 &
  384 x 288 x 48 &
  352 x 264 x 44 \\ \midrule
Phase/Slice oversampling    &   -    & 10/0 \%   & 10/0 \%   & 10/0 \%  & 10/0 \% \\ \midrule
TR &
  110.17 - 255.54 ms &
  2.37 ms &
  2.40 ms &
  2.40 ms &
  2.31 ms\\ \midrule
TE &
  \begin{tabular}[c]{@{}l@{}} 4.60 - 4.64 ms (In-Phase)\\ 2.30 ms (Opposed-Phase)\end{tabular} &
  1.00 ms &
  1.02 ms &
  1.02 ms &
  0.97 ms \\ \midrule
Flip angle                  & \ang{80}    & \ang{8}     & \ang{8}     & \ang{8}  & \ang{8}     \\ \midrule
Bandwidth                   &   -    & 920 Hz/Px & 930 Hz/Px & 930 Hz/Px & 950 Hz/Px\\ \midrule
GRAPPA factor               & None  & None         & None      & None    & None   \\ \midrule
Phase/Slice partial Fourier &   -    & Off/Off   & Off/Off   & Off/Off  & Off/Off \\ \midrule
Phase/Slice resolution      &   -    & 50/64 \%  & 50/64 \%  & 50/64 \% & 50/64 \% \\ \midrule
Fat saturation             &   -    & On      & On        & On   & On     \\ \midrule
Time per TP                 &   -    & 10.52 sec  & 12.80 sec  & 13.96 sec   & 11.36 sec \\ \toprule
\end{tabular}
}
\end{table*}






	
\begin{table*}
\centering
\caption{Effective resolutions and estimated acquisition times (per TP) of the dynamic and static datasets after performing different levels of artificial undersampling.}
\label{tab:3DDyn_ResTimeUnder} 
\resizebox{\textwidth}{!}{
\begin{tabular}{lcccccccc}
\toprule
                                                                          & \multicolumn{2}{c}{Protocol 1} & \multicolumn{2}{c}{Protocol 2} & \multicolumn{2}{c}{Protocol 3} & \multicolumn{2}{c}{Protocol 4} \\ \cline{2-9} 
 &
  \begin{tabular}[c]{@{}c@{}}Resolution\\ ($mm^3$)\end{tabular} &
  \begin{tabular}[c]{@{}c@{}}Acq. Time\\ ($sec$)\end{tabular} &
  \begin{tabular}[c]{@{}c@{}}Resolution\\ ($mm^3$)\end{tabular} &
  \begin{tabular}[c]{@{}c@{}}Acq. Time\\ ($sec$)\end{tabular} &
  \begin{tabular}[c]{@{}c@{}}Resolution\\ ($mm^3$)\end{tabular} &
  \begin{tabular}[c]{@{}c@{}}Acq. Time\\ ($sec$)\end{tabular} &
  \begin{tabular}[c]{@{}c@{}}Resolution\\ ($mm^3$)\end{tabular} &
  \begin{tabular}[c]{@{}c@{}}Acq. Time\\ ($sec$)\end{tabular} \\ \toprule
\begin{tabular}[c]{@{}l@{}}High Resolution \\ (Ground-truth)\end{tabular}& 0.90 x 0.90 x 4     & 8.76    & 0.90 x 0.90 x 4     & 10.68    & 0.90 x 0.90 x 4     & 11.76  & 1.00 x 1.00 x 4     & 9.38  \\
10\% of k-space                                                           & 2.70 x 2.70 x 4     & 0.88    & 2.70 x 2.70 x 4     & 1.07    & 2.70 x 2.70 x 4     & 1.18 & 3.00 x 3.00 x 4     & 0.94   \\
6.25\% of k-space                                                           & 3.60 x 3.60 x 4     & 0.55    & 3.60 x 3.60 x 4     & 0.67    & 3.60 x 3.60 x 4     & 0.74   & 4.00 x 4.00 x 4     & 0.59  \\
4\% of k-space                                                         & 4.50 x 4.50 x 4     & 0.35    & 4.47 x 4.47 x 4     & 0.43    & 4.47 x 4.47 x 4     & 0.47  & 4.99 x 4.99 x 4     & 0.38  \\ 
\toprule
\end{tabular}
}
\end{table*}
\subsubsection{Dynamic Data Generation}
\label{sec:chaosdyn}
Since large dynamic MRI datasets that would be required for training are not available publicly, an artificial dynamic dataset was created. This was achieved by applying random elastic deformation of TorchIO~\citep{perez2021torchio} on the volumes from the CHAOS dataset. Random displacement fields were generated using Torchio's random elastic deformation with five control points, 5-20-20 mm of maximum displacements along x-y-z dimensions, respectively, and two locked borders. The displacement files were then applied to the volumes of the CHAOS dataset using cubic B-spline interpolation, considering them as TP0, to generate artificial TP1. Then, a new set of random displacement fields with the same parameters were generated and applied on TP1 to generate TP2. In this manner, 24 artificial time-points (TP1 - TP24) were generated for each of the volumes present in the original dataset. The displacement field tries to imitate the movement induced by breathing during a dynamic acquisition. The displacement field was set to expand and/or contract more in anterosuperior (front-back) and superoinferior (up-down) but less in lateral (left-right) direction. However, this manner of generating artificial breathing motion is not equivalent to physiological motion. It is to be noted that the goal of using this kind of artificial motion was to create a dataset from which a network can learn the pseudo-temporal relationship between two subsequent time-points. This process results in an artificially created dynamic dataset - CHAOS dynamic, comprising 25 time-points in total for each volume. 

\subsubsection{Undersampling} 
The training data - the original CHAOS dataset and the artificially created CHAOS dynamic dataset, as well the testing data (3D dynamic scans) were artificially undersampled in-plane using MRUnder~\citep{soumick_chatterjee_2020_3901455,chatterjee2021reconresnet}\footnote{MRUnder on Github: \url{https://github.com/soumickmj/MRUnder}} by taking only 10\%, 6.25\%, and 4\% of the centre k-space, resulting in MR acceleration factors of 3, 4, and 5, respectively (considering undersampling only in the phase-encoding
direction). Considering the actual amount of data used during SR-reconstruction, this results in theoretical acceleration factors of 10, 16, and 25, respectively. The effective resolutions and estimated acquisition times for each of the dynamic test datasets are calculated using Eq.~\ref{eq:acqtime} and shown in  Table~\ref{tab:3DDyn_ResTimeUnder}

\begin{equation}\label{eq:acqtime}
 T_{acq} =PE_{n} \times TR \times S_{m}
\end{equation}

where $T_{acq}$ is the estimated acquisition time, given the number of phase-encoding lines $PE_{n}$, the repetition time $TR$, and the number of slices acquired $S_{m}$~\citep{sarasaen2021fine}. During the calculation of Table~\ref{tab:3DDyn_ResTimeUnder}, phase/slice resolution and phase/slice oversampling (Table~\ref{tab:MRIparams_CHAOS_3DDyn}) were also taken into consideration while calculating $PE_{n}$ and $S_{m}$.




\subsection{Implementation, Training, and Inference} \label{sec:implement}
\begin{figure}[!htbp]
\centering
\includegraphics[width=0.48\textwidth]{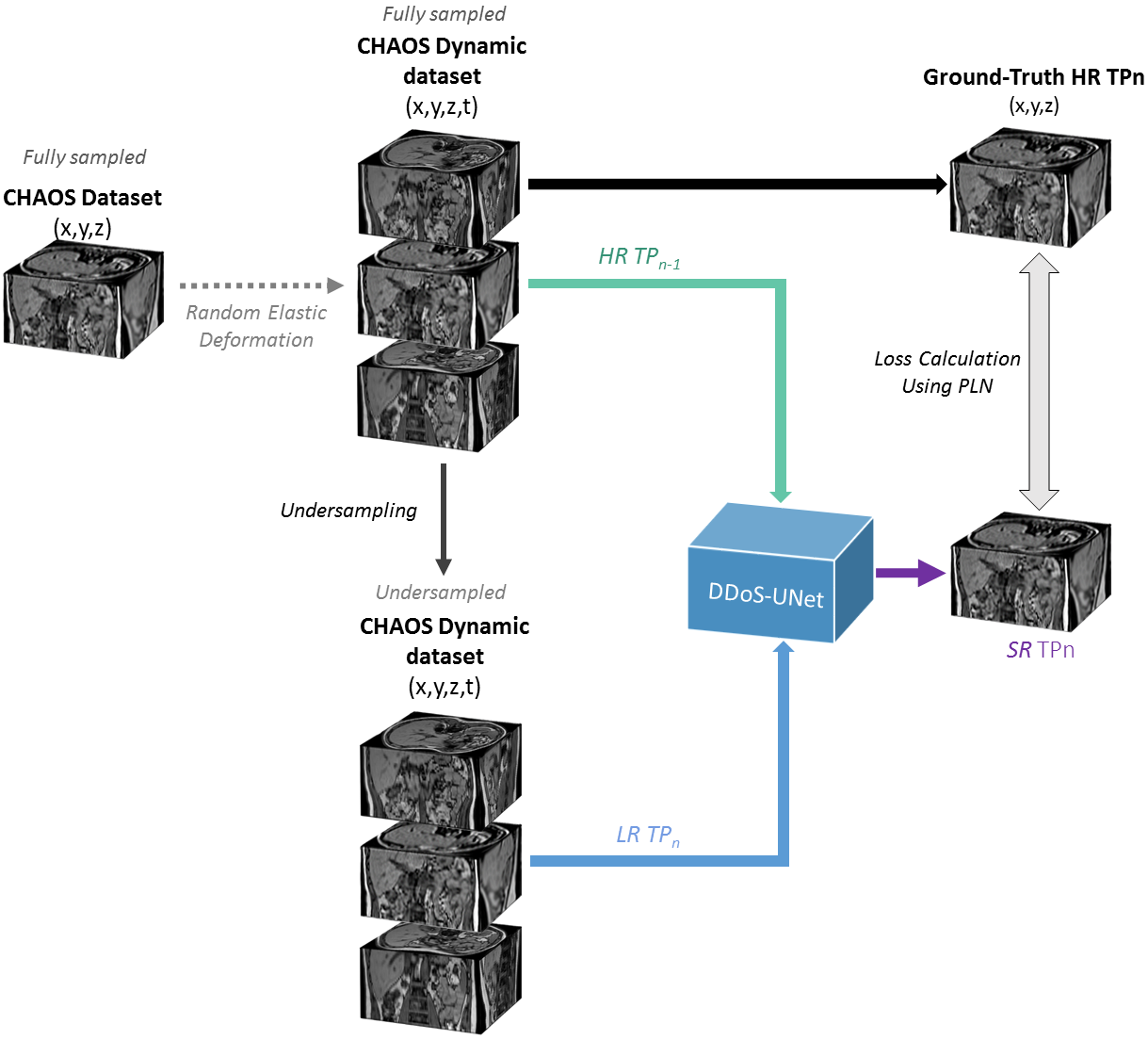}
\captionsetup{justification=centering}
\caption{Method Overview: Training. Initially, random elastic deformation is applied on the CHAOS dataset (fully sampled) to generate the artificial CHAOS dynamic dataset. Then the CHAOS dynamic dataset was undersampled to generate the final training dataset. Then the model is trained by providing low-resolution (undersampled) current time-point ($LR TP_{n}$) along with the high-resolution (fully sampled) previous time-point ($HR TP_{n-1}$) as input and the output is compared against the ground-truth high-resolution current time-point ($HR TP_{n}$).}
\label{fig:training}
\end{figure}

\begin{figure}[!htbp]
\centering
\includegraphics[width=0.48\textwidth]{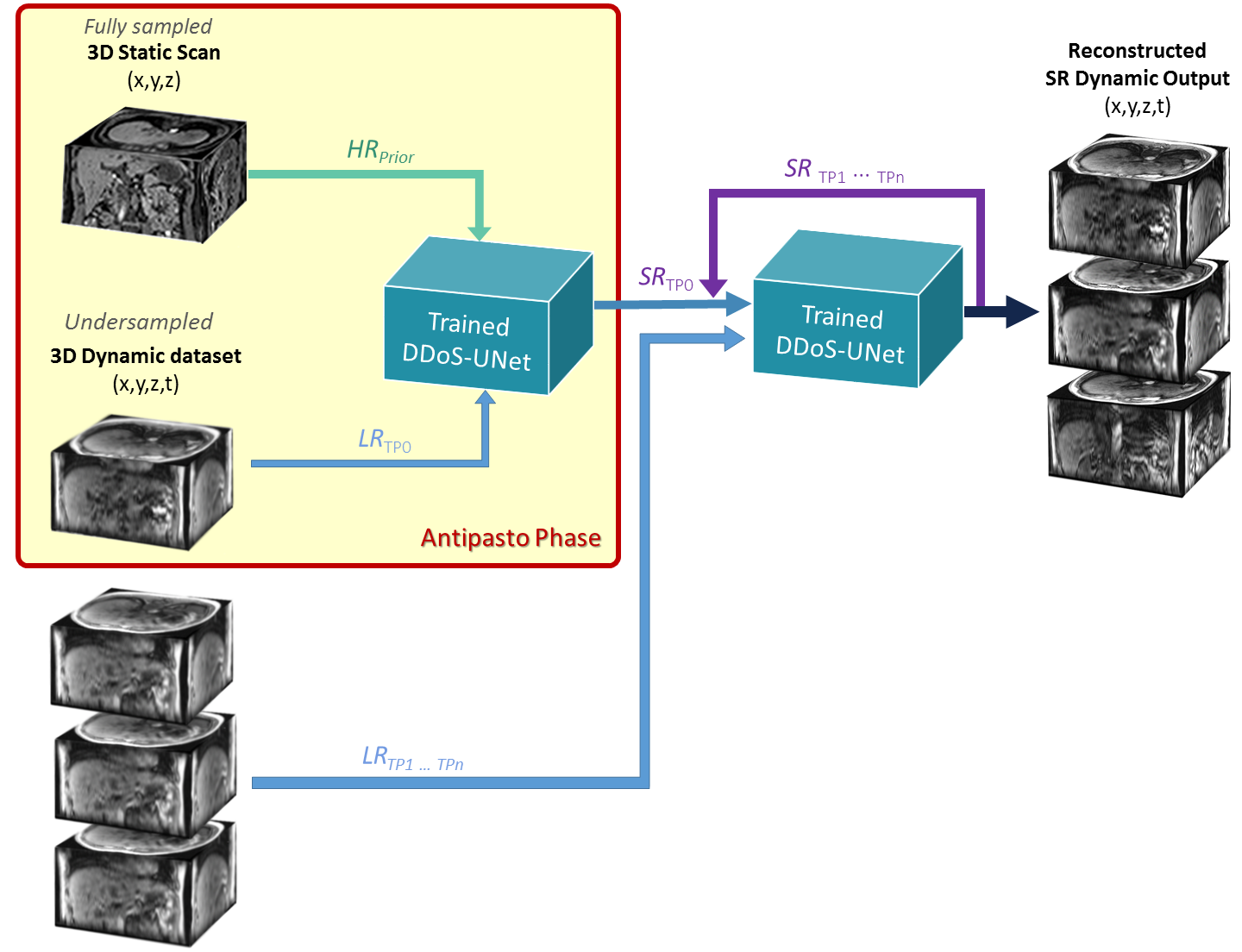}
\captionsetup{justification=centering}
\caption{Method Overview: Inference. 3D static subject-specific planning scan (fully sampled) is supplied as the high-resolution prior image ($HR_{Prior}$) along with the first low-resolution (undersampled) time-point ($LR_{TP0}$) of the 3D dynamic dataset are supplied as input to the trained DDoS-UNet model, and the model super-resolves $LR_{TP0}$ to obtain $SR_{TP0}$. This initial phase is called as the "Antipasto" phase. $LR_{TP0}$ is then supplied as input together with the next low-resolution time-point $LR_{TP1}$ to the same trained DDoS-UNet model to obtain $SR_{TP1}$. This process is continued recursively until all the time-points of the low-resolution (undersampled) 3D dynamic are super-resolved, by supplying pairs of $SR_{TPn-1}$ and $LR_{TPn}$ to obtain each of the $SR_{TPn}$.}   
\label{fig:inference}
\end{figure}

The proposed model was trained on 3D volumes from the artificially created dynamic version of a publicly available benchmark dataset, as summarised in Fig.~\ref{fig:training}. Fig.~\ref{fig:inference} shows an overview of the inference steps. The inference process starts with the Antipasto phase - by supplying the high-resolution patient-specific static scan as a prior image on the first channel of the network (as $\hat{HR}_{TP(n-1)}$ is not yet available), and by supplying $LR_{TP(0)}$ on the second channel of the network. 


It is to be noted that the static scan has the same resolution, contrast and volume coverage as the high-resolution ground-truth dynamic scan. However, to keep the testing environment similar to the real-life scenario and keep a fast speed of inference, the static and dynamic datasets were not co-registered, as registration is typically time-consuming. After this, the network super-resolves $LR_{TP(0)}$ to $\hat{HR}_{TP(0)}$. Now for the next time-point, $\hat{HR}_{TP(0)}$ and $LR_{TP(1)}$ are supplied as input to the network and the network provides $\hat{HR}_{TP(1)}$ as output.

The implementation was done using PyTorch~\citep{NEURIPS2019_9015}, and training-inference were performed using Nvidia Tesla V100 GPUs. Following the hypothesis of using batch size one to be able to learn an exact mapping function between the specific pair of low-high-resolution images~\citep{chatterjee2021reconresnet}, batch size during training and inference in this research was also set to one. The loss during training was calculated using perceptual loss~\citep{johnson2016perceptual}, with the help of a perceptual loss network~\cite{chatterjee2020ds6}, and was minimised using the Adam optimiser with a learning rate of 1e-4 for 100 epochs. The code of the implementation is available on GitHub\footnote{DDoS on Github: \url{https://github.com/soumickmj/DDoS}}.

\subsubsection{Perceptual Loss} \label{sec:ploss}
Similar to the previous work~\citep{sarasaen2021fine}, perceptual loss~\citep{johnson2016perceptual} was employed to compute the loss during training. For the same, the initial three blocks of the frozen pre-trained (on 7T MRA scans, for the task of vessel segmentation) UNet MSS model was used as the perceptual loss network (PLN)~\citep{chatterjee2020ds6}. The job of this PLN is to extract minute features of different abstraction levels at the different levels of the PLN, from the super-resolved volumes and their corresponding ground-truths. The features generated from the super-resolved output and the ground-truth were compared against each other using mean absolute error (L1 loss). Finally, all these l1 losses were added together and backpropagated. 


\subsection{Evaluation Criteria} \label{sec:evalCriteria}
The quality of super-resolution was evaluated quantitatively with the help of the structural similarity index (SSIM)\citep{wang2004imageSSIM}, the peak signal-to-noise ratio (PSNR), and the normalised root mean squared error (NRMSE). The perceptual quality of the output was evaluated with the help of SSIM, which compares luminance, contract, and structure terms between two given images $x$ and $y$, which for this research represent the output and ground-truth, respectively, using the following formula:
\begin{equation}
    SSIM (x,y) = \frac{(2\mu_x\mu_y+c_1)(2\sigma_{xy}+c_2)}{(\mu_x^2+\mu_y^2+c_1)(\sigma_x^2+\sigma_y^2+c_2)}
    \label{eqn:SSIM} 
\end{equation}
where $\mu_x, \mu_y, \sigma_x, \sigma_y$ and $\sigma_{xy}$ are the local means, standard deviations, and cross-covariance for images $x$ and $y$, respectively. $c_{1}=(k_{1}L)^{2}$ and $c_{2}=(k_{2}L)^{2}$, where $L$ is the dynamic range of the pixel-values, $k_{1}=0.01$ and $k_{2}=0.03$.
Moreover, the quality of the super-resolution was measured statistically with the help of PSNR and NRMSE, both of which are calculated using the mean-square error (MSE) between $x$ and $y$ as:
\begin{equation}
    PSNR = 10 \log_{10} \left(\frac{R^2}{MSE}\right)
    \label{eqn:PSNR} 
\end{equation}
where $R$ is the maximum fluctuation in the input image, and
\begin{equation}
     NRMSE = \frac{\sqrt{MSE} * \sqrt{N}}{|| y ||}
 \end{equation}
 where $|| \cdot ||$ denotes the Frobenius norm, $N$ is the number of elements in the data, and $y$ is the ground-truth.
 
The statistical significance of the differences in the quantitative metrics for the proposed method against the other baselines was computed using the Mann–Whitney U test. Apart from quantitative evaluations, the results were also compared qualitatively. 


\section{Results}\label{sec:Results}

The performance of the DDoS-UNet was compared for three different levels of undersampling: 10\%, 6.25\%, and 4\% of the centre k-space, against the low-resolution input, traditional trilinear-interpolation, Fourier interpolated input (zero-padded k-space), and finally against two different baseline deep learning models: two UNet models identical to the DDoS-UNet except for the initial layer (unlike DDoS-UNet, these UNets received one input) - one of them trained on the original CHAOS dataset (T1-dual images, in- and opposed phase)~\citep{kavur2021chaos}, and the other one was trained using artificial dynamic CHAOS (see Sec.~\ref{sec:chaosdyn}). The training dataset of the second UNet was identical to the training dataset of the DDoS-UNet. The models were evaluated on real dynamic datasets of five subjects, each consisting of 25 time-points (details in Sec.~\ref{sec:data}). The inference process for the DDoS-UNet was started with the patient-specific prior high-resolution static scan and first low-resolution time-point as input and then continued by supplying the previous super-resolved time-point with the current low-resolution time-point to super-resolve the current time-point (as explained in Sec.~\ref{sec:ddos}). 

\begin{figure*}
\centering
\includegraphics[width=\textwidth]{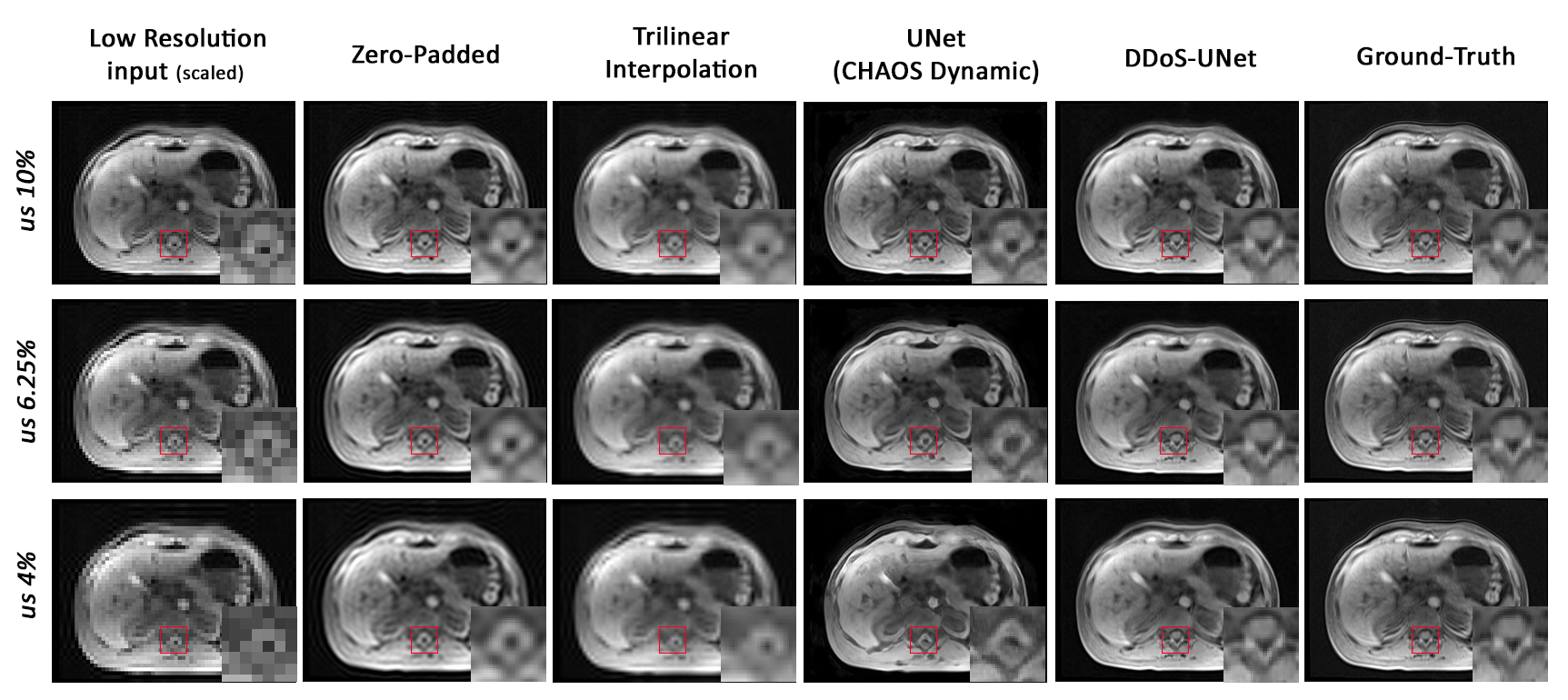}
\captionsetup{justification=centering}
\caption{Comparative results of low resolution (10\%, 6.25\% and 4\% of k-space) 3D Dynamic data of the same slice. From left to right: low resolution images (scaled-up, nearest-neighbour interpolation), Interpolated input (Trilinear), Zero-padded reconstruction, Output of UNet trained on CHAOS dynamic dataset, Output of DDoS-UNet and ground-truth images.}
\label{fig:3DDyn_Qualitativecompare}
\end{figure*}


\begin{figure*}
\centering
\includegraphics[width=0.85\textwidth]{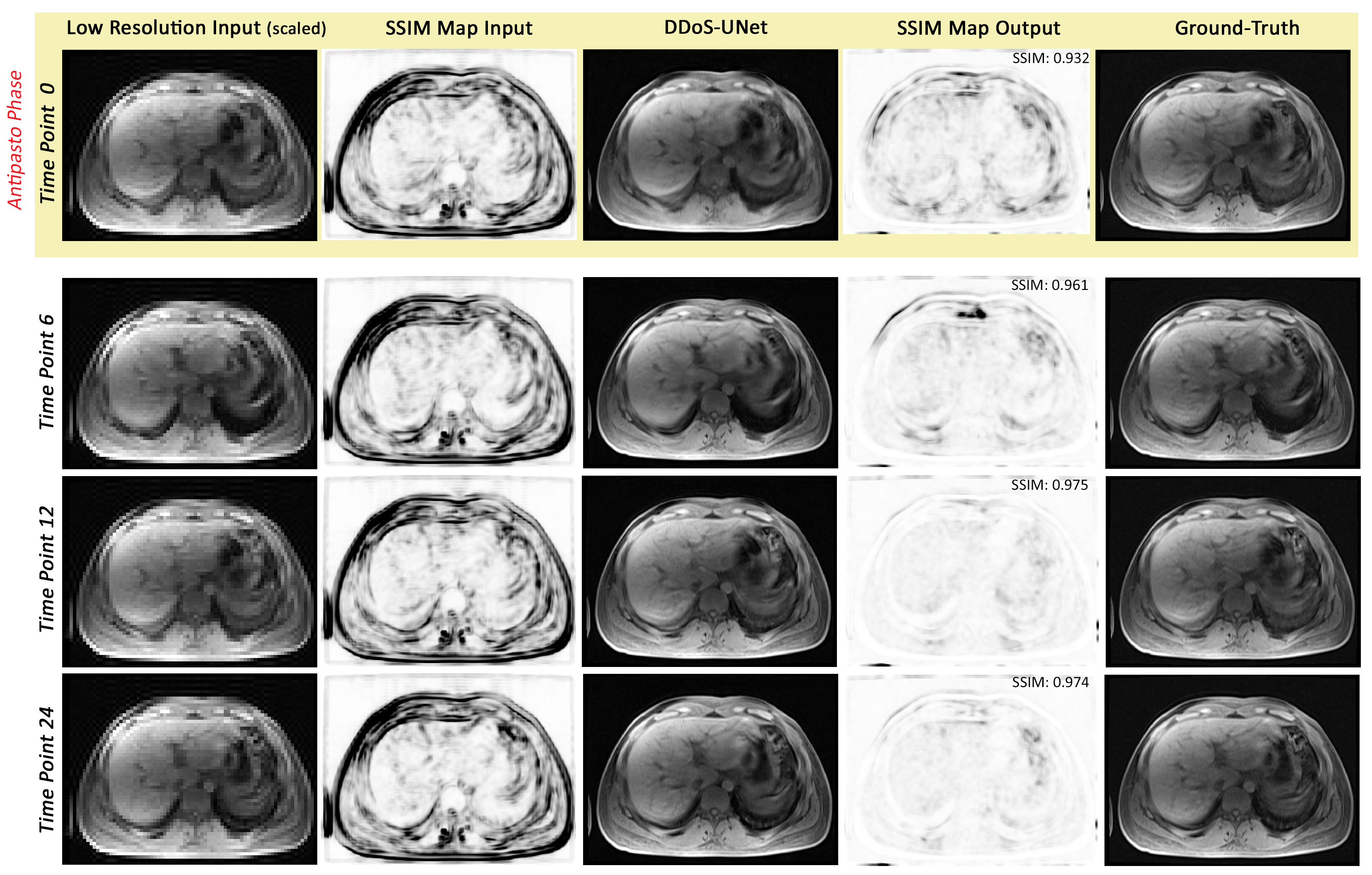}
\captionsetup{justification=centering}
\caption{An example comparison of the low resolution input of the 4\% of k-space with the super-resolution (SR) result of the DDoS-UNet over four different time points, compared against the high resolution ground-truth using SSIM maps.}
\label{fig:input_output_SSIMmaps}
\end{figure*}

\begin{figure*}
\centering
\includegraphics[width=0.90\textwidth]{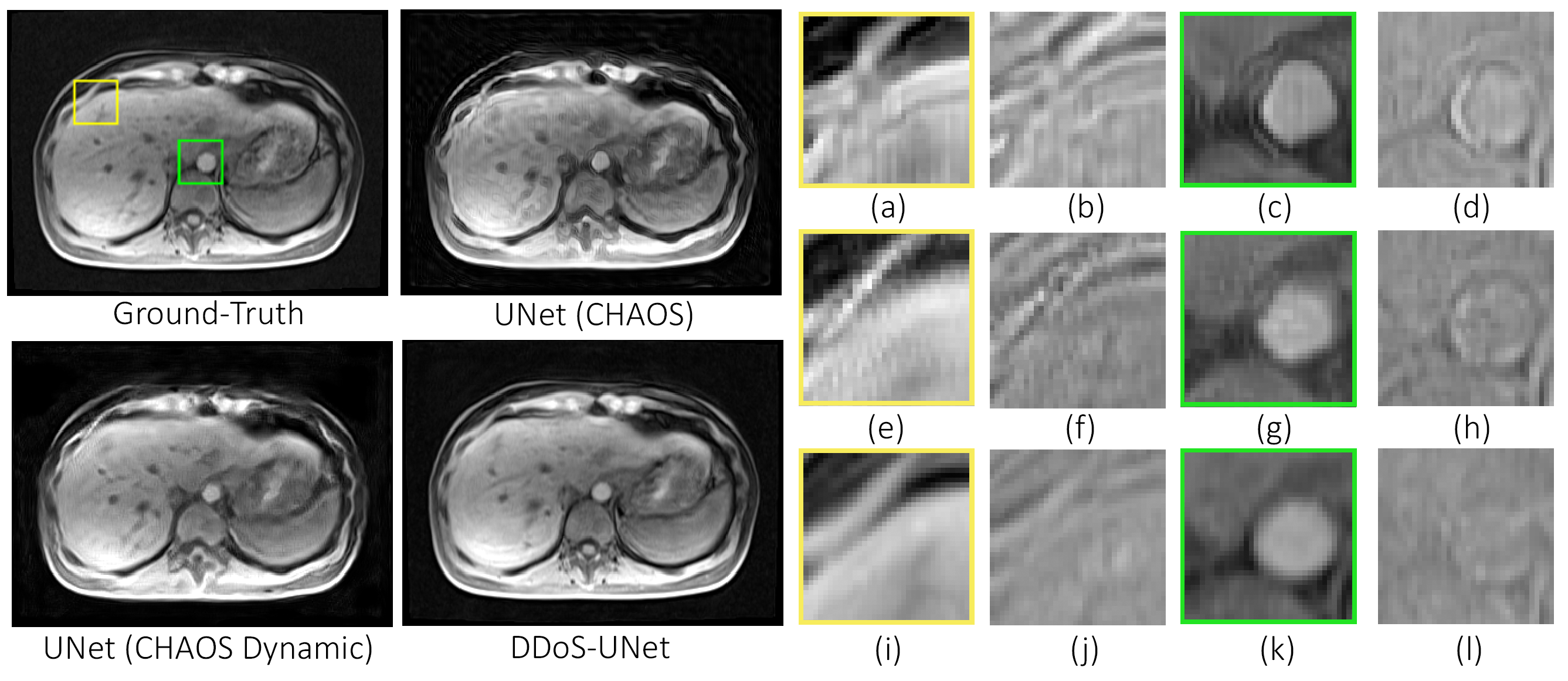}
\captionsetup{justification=centering}
\caption{An example of reconstructed results from UNet baselines and DDoS-UNet, compared against its ground-truth (GT) for low resolution images from 4\% of k-space. From left to right, upper to lower: ground-truth,  SR result of the UNet baseline (UNet CHAOS), SR result of the UNet baseline trained on CHAOS dynamic (UNet CHAOS Dynamic) and SR result of the DDoS-UNet. For the yellow ROI, (a-b): UNet CHAOS and the difference image from GT, (e-f): SR result of UNet CHAOS Dynamic and (i-j): SR result of DDos-UNet and the difference image from GT. The images on the right part are identical examples for the green ROI.}

\label{fig:roiUs4Subtract}
\end{figure*}

\begin{table*}
\centering
\caption{The average and the standard deviation of SSIM, PSNR, and NRMSE. The table shows the results for different resolutions.}
\label{tab:SSIMPSNRdiffSD}
\resizebox{\textwidth}{!}{%
\begin{tabular}{llllllllll}
\hline
\multicolumn{1}{c}{\multirow{2}{*}{Data}} & \multicolumn{3}{c}{10\% of k-space}      & \multicolumn{3}{c}{6.25\% of k-space}    & \multicolumn{3}{c}{4\% of k-space}       \\ \cline{2-10} 
\multicolumn{1}{c}{} &
  \multicolumn{1}{c}{SSIM} &
  \multicolumn{1}{c}{PSNR} &
  \multicolumn{1}{c}{NRMSE} &
  \multicolumn{1}{c}{SSIM} &
  \multicolumn{1}{c}{PSNR} &
  \multicolumn{1}{c}{NRMSE} &
  \multicolumn{1}{c}{SSIM} &
  \multicolumn{1}{c}{PSNR} &
  \multicolumn{1}{c}{NRMSE} \\ \hline
Trilinear Interpolation                   & 0.872±0.014 & 28.631±1.364 & 0.192±0.023 & 0.821±0.017 & 26.770±1.226 & 0.238±0.024 & 0.765±0.022 & 25.248±1.298 & 0.283±0.025 \\
Zero-padded                               & 0.949±0.013 & 36.138±1.753 & 0.082±0.016 & 0.910±0.018 & 29.761±1.640 & 0.124±0.019 & 0.863±0.021 & 32.520±1.508 & 0.170±0.025 \\
UNet (CHAOS)                              & 0.967±0.006 & 38.359±1.580 & 0.021±0.004 & 0.944±0.010 & 35.623±1.552 & 0.029±0.005 & 0.916±0.015 & 32.658±1.598 & 0.041±0.007 \\
UNet (CHAOS Dynamic)                      & 0.959±0.012 & 37.376±1.275 & 0.024±0.003 & 0.941±0.012 & 35.113±1.566 & 0.031±0.006 & 0.914±0.012 & 33.620±1.035 & 0.036±0.004 \\
\textbf{DDoS-UNet} &
  \textbf{0.980±0.006} &
  \textbf{41.824±2.070} &
  \textbf{0.014±0.003} &
  \textbf{0.967±0.011} &
  \textbf{39.494±2.121} &
  \textbf{0.019±0.005} &
  \textbf{0.951±0.017} &
  \textbf{37.557±2.179} &
  \textbf{0.024±0.006} \\ \hline
\end{tabular}%
}
\end{table*}
\begin{figure*}
\centering
\includegraphics[width=\textwidth]{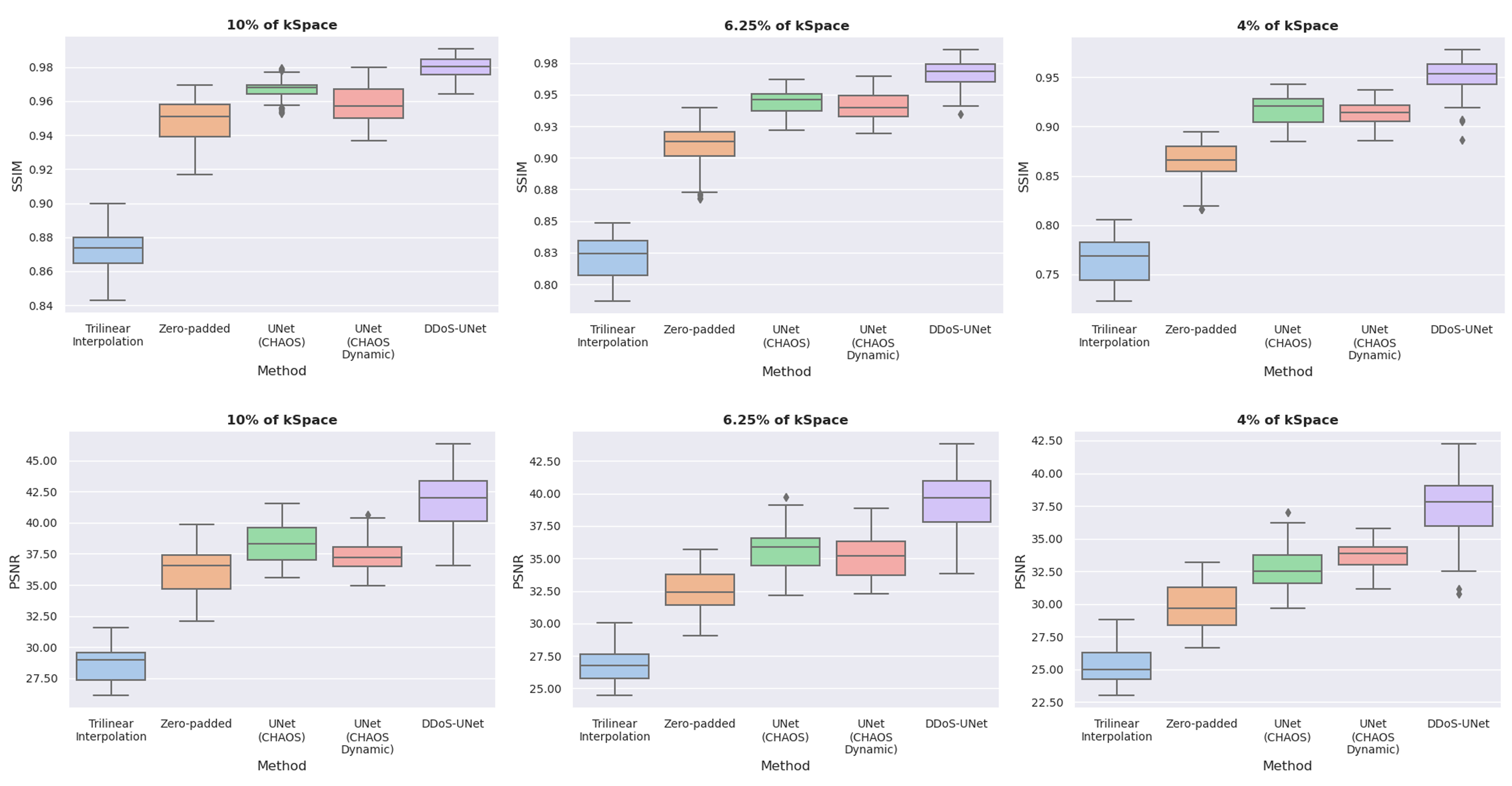}
\captionsetup{justification=centering}
\caption{Quantitative comparison of different methods using SSIM and PSNR - for all subjects and time-points combined, for different levels of undersampling.}
\label{fig:Box_SSIMPSNR}
\end{figure*}


Fig.~\ref{fig:3DDyn_Qualitativecompare} shows a qualitative comparison of the results obtained by the different methods for different levels of undersampling. It can be observed that the proposed DDoS-UNet managed to restore finer details better than the other methods. Moreover,  both the baseline UNet models show better anatomical structures than the zero-padded reconstructions. Furthermore, the comparison with the help of SSIM maps between the input (low-resolution images) and output (super-resolved images) of the DDoS-UNet are shown in Fig.~\ref{fig:input_output_SSIMmaps}. It reveals that the reconstruction quality of the initial time-point is not very good, but the network manages to recover from the initial struggle during the Antipasto phase, and manages to reconstruct the subsequent time-points much better and consistently over all the time-points. This can be attributed to the fact that the static and dynamic scans are acquired in two different sessions, and they are not co-registered. Finally, Fig.~\ref{fig:roiUs4Subtract} shows the qualitative comparisons of the different methods for two regions-of-interest (ROI). It shows the proposed DDoS-UNet framework results in better reconstruction performance than the baseline UNet models. Between the baseline UNets, UNet trained on CHAOS dynamic dataset managed to recover finer anatomical details better than the UNet model trained on CHAOS dataset.

\begin{figure*}
\centering
\includegraphics[width=\textwidth]{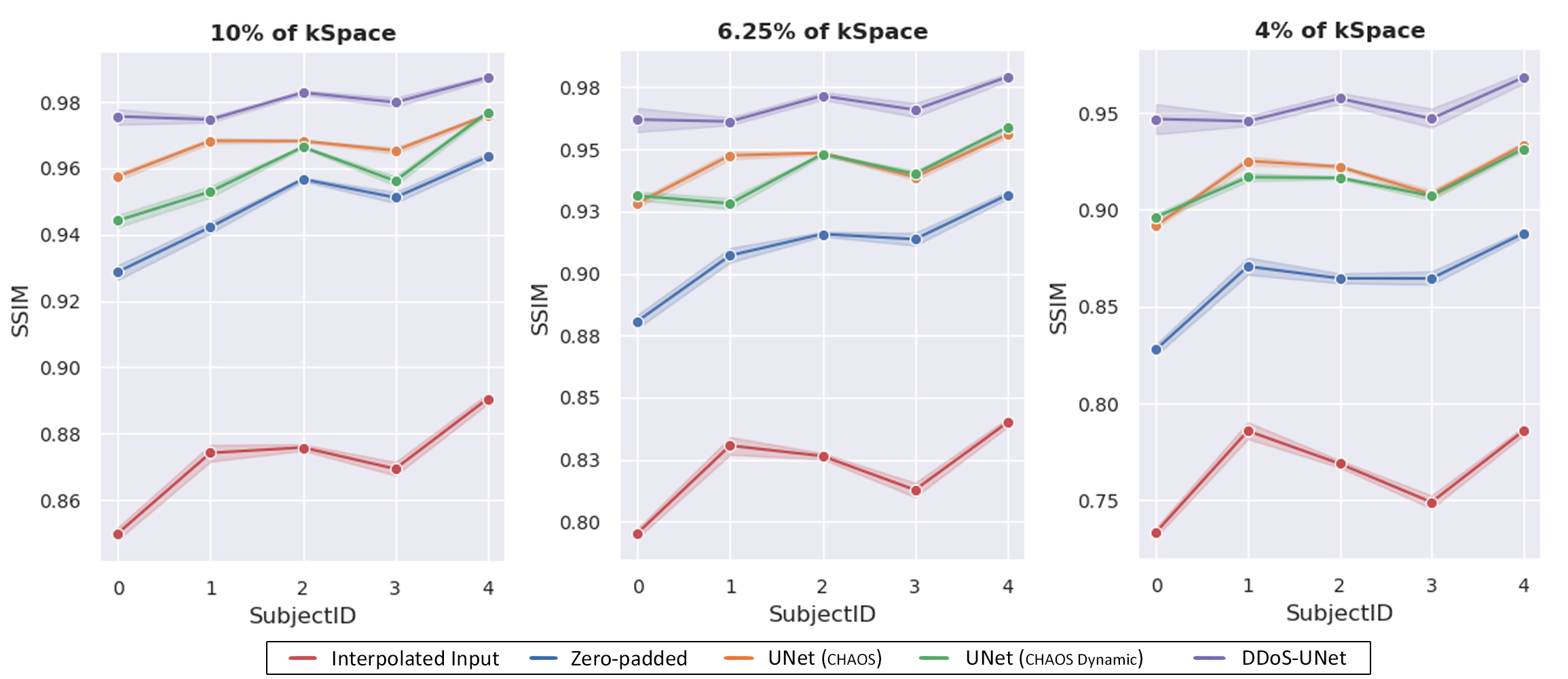}
\captionsetup{justification=centering}
\caption{Line plot showing the mean and 95\% confidence interval of the resultant SSIM values over the different time-points (excluding the initial one, the Antipasto phase) for each subject. The red, blue, orange, green, and violet lines represent the reconstruction results of trilinear interpolation, zero-padding (sinc interpolation), UNet trained on CHAOS dataset, UNet trained on CHAOS Dynamic dataset, and DDoS-UNet, respectively.}
\label{fig:SSIM_Sub}
\end{figure*}

Table~\ref{tab:SSIMPSNRdiffSD} presents the quantitative results for all the methods. It can be observed that both the baseline UNet models (trained on the original CHAOS dataset and on the CHAOS dynamic dataset) outperformed the non-DL baselines: trilinear interpolation and zero-padded reconstruction (sinc interpolation), and the proposed DDoS-UNet method outperformed all the baselines for all three undersampling factors in all three metrics with statistical significance (p-values always less than $0.001$). It can be further observed that the UNet trained on the original CHAOS dataset outperformed the UNet trained on the CHAOS dynamic dataset. Fig.~\ref{fig:Box_SSIMPSNR} shows the resultant SSIM and PSNR values over all subjects and time-points by means of box plots. It can be observed that the improvements obtained by the proposed method increase with the increase in the undersampling factor. Fig.~\ref{fig:SSIM_TP} portrays the SSIM values over the different time-points averaged for all five subjects. It can be seen that after the initial time-point TP0 (Antipasto phase), the proposed DDoS-UNet achieved consistently better SSIM values compared to all the other methods. Finally, Fig.~\ref{fig:SSIM_Sub} shows the average SSIM values over the different time-points (excluding the Antipasto phase) for each subject. The median values over TP1 to TP24 for each of the subjects resulted in SSIM values in the range $0.988$ to $0.975$, $0.980$ to $0.960$, and $0.970$ to $0.945$, for 10\%, 6.25\%, and 4\% of k-space, respectively. Fig.~\ref{fig:SSIM_TP}~and~\ref{fig:SSIM_Sub} show that the proposed DDoS-UNet is able to reconstruct different protocols and subjects efficiently while being stable over different time-points.  

\section{Discussion}\label{sec:Discussion}
This paper presents the \textbf{D}ynamic \textbf{D}ual-channel \textbf{o}f \textbf{S}uper-resolution using \textbf{UNet} (DDoS-UNet) framework and shows its applicability for reconstructing low-resolution (undersampled) dynamic MRIs up to a theoretical acceleration factor of 25. The quantitative and qualitative results demonstrate the superiority of the proposed method. 

The UNet model trained on the original CHAOS dataset performed better quantitatively than the UNet model trained on the CHAOS dynamic dataset, even though the latter had 25 times more volumes (24 artificially created time-points on top of the original one). This can be attributed to the quality of the CHAOS dynamic dataset. Due to the repeated applications of the random elastic deformation on the original dataset, which includes interpolation, the sharpness of the later time-points decreased caused by the accumulated interpolation errors. This might have also negatively impacted the results of the DDoS-UNet. Improving the quality of the artificial dynamic dataset might improve the performance of both of these models. It is worth mentioning, however, that for the highest undersampling factor (4\% of the k-space), UNet trained on CHAOS dynamic dataset resulted in better PSNR than UNet trained on CHAOS dataset, and also visual comparison (Fig.~\ref{fig:roiUs4Subtract}) revealed that the UNet trained on CHAOS dynamic managed to restore finer anatomical details better. 

\begin{figure*}
\centering
\includegraphics[width=\textwidth]{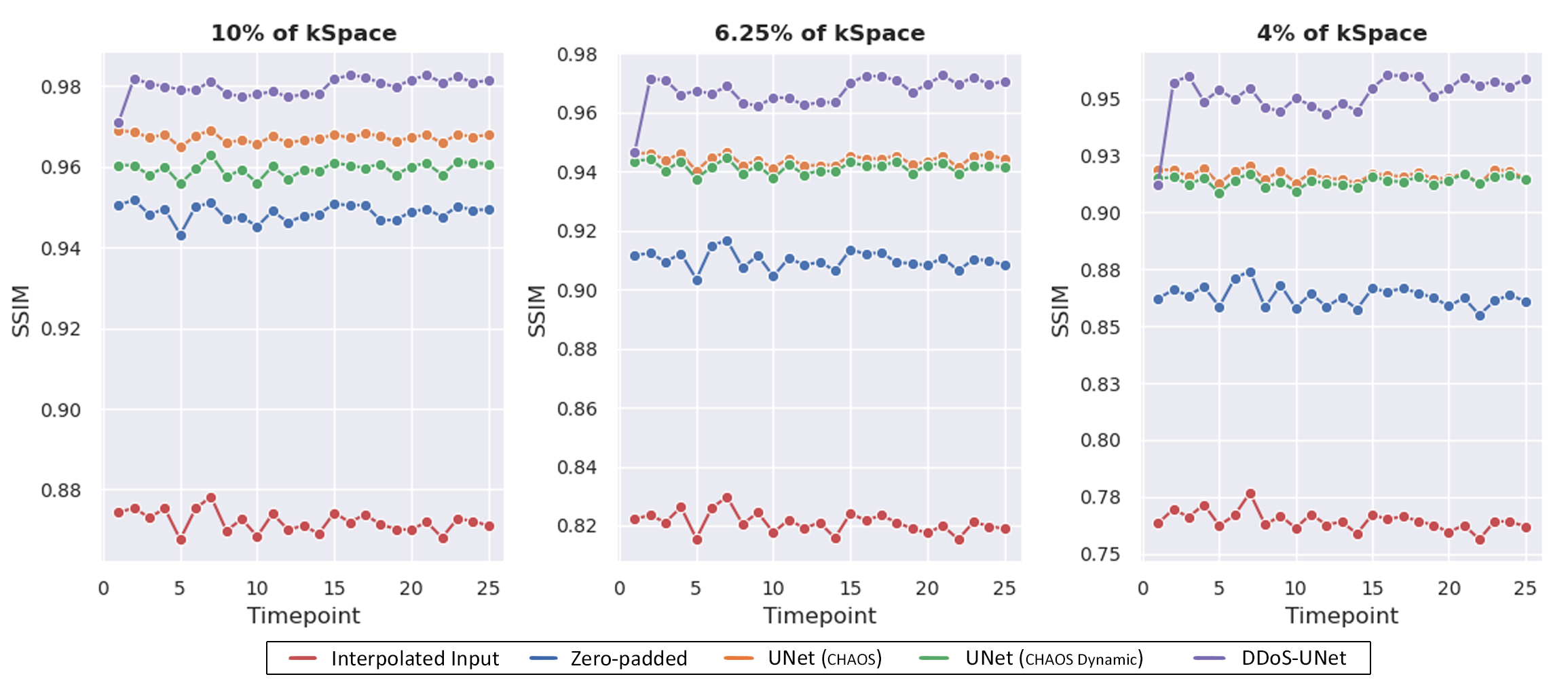}
\captionsetup{justification=centering}
\caption{Line plot showing the average SSIM values over each subject for for all time-points, for different levels of undersampling. Initial drop can be observed for the first time point for DDoS-UNet, which is referred here as the Antipasto phase, then the network peforms with stability for the rest of the time-points.}
\label{fig:SSIM_TP}
\end{figure*}

A final observation can be made regarding the results of the DDoS-UNet for the different time-points. The result of the initial time-point was considerably worse compared to the rest of the other time-points (similar or better than the UNets, and always better than the non-DL baselines), as can be seen in Figures~\ref{fig:SSIM_TP}~and~\ref{fig:input_output_SSIMmaps}. This initial time-point was reconstructed by supplying the high-resolution subject-specific static scan as the prior image, referred here as the Antipasto phase, whereas the rest of the time-points were reconstructed by supplying the super-resolved previous time-point as the prior image. The static scan has a big temporal difference from the first time-point of the dynamic scan as they were acquired in different sessions, while the subsequent time-points of the dynamic scan were closer in time. The network faces difficulties reconstructing the initial time-point, but then recovers from it after super-resolving the first one and then maintaining its performance steadily for all subsequent time-points. This also supports the hypothesis that the DDoS-UNet learnt both spatial and temporal relationships, as shown in Eqs.~\ref{eqn:DDoS}~and~\ref{eqn:DDoS_param} in Sec.~\ref{sec:ddos}. 

The reconstruction (inference) time using the proposed DDoS-UNet was approximately 0.36 secs for each time-point (9 secs for 25 TPs) while reconstructing using an Nvidia Tesla V100 GPU. Fast reconstruction time, coupled with the high speed of acquisition (shown in Table~\ref{tab:3DDyn_ResTimeUnder}), this method shows the potential to acquire-reconstruct each time-point of a 3D dynamic acquisition within 0.71 secs (for 4\% of k-space with Protocol 1) - making it a potential candidate for near real-time MR acquisitions. The acquisition time can be further reduced using techniques such as parallel imaging, as shown in the earlier work~\citep{sarasaen2021fine}. The focus of this paper is on abdominal imaging; however, this method might also be used for other types of dynamic imaging, e.g. cardiac imaging.


\section{Conclusion and Future Work}\label{sec:Conclusion}
This research proposes the DDoS-UNet model to perform 3D volumetric super-resolution of low-resolution dynamic MRIs by using a subject-specific high-resolution prior planning scan and exploiting the spatio-temporal relationship present in the dynamic MRI. The proposed network was trained using an artificially created dynamic dataset from the CHAOS abdominal benchmark dataset and then was tested using dynamic MRIs comprising of 25 time-points. It was observed that even though the network was trained using a dataset with MRI acquisition parameters very different from the test set, the network was able to super-resolve the given input images with high accuracy - even for high undersampling factors. The proposed method resulted in 0.951±0.017 SSIM while super-resolving the highest undersampling experimented in this research (i.e. 4\% centre k-space), whereas the baseline UNet (model without supplying the super-resolved previous time-point as prior information) resulted in 0.916±0.015. The results show that the proposed network managed to mitigate the spatio-temporal problem of dynamic MRI by performing spatial super-resolution with the help of the temporal relationship present in the data without compromising the acquisition speed. Given the reconstruction speed of the proposed approach, this can be a candidate for near real-time dynamic acquisition scenarios, such as interventional MRI. 

The proposed approach employs a multi-channel approach to supply the prior image (initially, the high-resolution static scan, then the super-resolved volumes). However, other approaches such as dual-branch have also been proposed~\citep{chatterjee2020retrospective}, which might also be used to supply such prior images to the network. Such an architecture can deal with the prior image and the low-resolution image differently (i.e. different weights applied on each), whereas the current initial layer of the network treats them equally and merges them as an internal representation in the initial layer. Moreover, DDoS-UNet is interesting in interventional setup. During interventions, devices such as catheters are used, which were not present in the training set. The authors plan to extend the current research by evaluating the proposed model's reconstruction performance for such devices. 


\section*{Acknowledgement}

This work was conducted within the context of the International Graduate School MEMoRIAL at Otto von Guericke
University (OVGU) Magdeburg, Germany, kindly supported by the European Structural and Investment Funds (ESF) under the
programme "Sachsen-Anhalt WISSENSCHAFT Internationalisierung" (project no. ZS/2016/08/80646).


\bibliography{mybibfile}
\end{document}